# Imaging Hydrodynamic Electrons Flowing Without Landauer-Sharvin Resistance

C. Kumar[1†], J. Birkbeck[1†], J.A. Sulpizio[1†], D. Perello[2,3], T. Taniguchi[4], K. Watanabe[4], O. Reuven[1], T. Scaffidi[5], Ady Stern[1], A.K. Geim[2,3], and S. Ilani[1*]

[1] *Department of Condensed Matter Physics, Weizmann Institute of Science, Rehovot 76100, Israel.*
[2] *School of Physics & Astronomy, University of Manchester, Manchester M13 9PL, United Kingdom.*
[3] *National Graphene Institute, University of Manchester, Manchester M13 9PL, United Kingdom.*
[4] *National Institute for Materials Science, 1-1 Namiki, Tsukuba, 305-0044 Japan.*
[5] *Department of Physics, University of Toronto, Ontario M5S 1A7, Canada.*
† These authors contributed equally to the work.
* Correspondence to: shahal.ilani@weizmann.ac.il

**Electrical resistance usually originates from lattice imperfections. However, even a perfect lattice has a fundamental resistance limit, given by the Landauer[1] conductance caused by a finite number of propagating electron modes. This resistance, shown by Sharvin[2] to appear at the contacts of electronic devices, sets the ultimate conduction limit of non-interacting electrons. Recent years have seen growing evidence of hydrodynamic electronic phenomena[3–18], prompting recent theories[19,20] to ask whether an electronic fluid can radically break the fundamental Landauer-Sharvin limit. Here, we use single-electron-transistor imaging of electronic flow in high-mobility graphene Corbino disk devices to answer this question. First, by imaging ballistic flows at liquid-helium temperatures, we observe a Landauer-Sharvin resistance that does not appear at the contacts but is instead distributed throughout the bulk. This underpins the phase-space origin of this resistance - as emerging from spatial gradients in the number of conduction modes. At elevated temperatures, by identifying and accounting for electron-phonon scattering, we reveal the details of the purely hydrodynamic flow. Strikingly, we find that electron hydrodynamics eliminates the bulk Landauer-Sharvin resistance. Finally, by imaging spiraling magneto-hydrodynamic Corbino flows, we reveal the key emergent length-scale predicted by hydrodynamic theories – the Gurzhi length. These observations demonstrate that electronic fluids can dramatically transcend the fundamental limitations of ballistic electrons, with important implications for fundamental science and future technologies.**

Electrical resistance is synonymous with the back-scattering of electrons. When an electron collides with an impurity, a phonon, or a rough device edge, it loses some of its momentum to the lattice, generating resistance. Therefore, it is very surprising that even when all these back-scattering sources are eliminated, an electronic device still has a non-zero resistance. As shown by Landauer[1], such resistance results from the finite conduction capacity of a channel, given by the number of its conduction modes multiplied by $e^2/h$ ($e$ is the electronic charge and $h$ is Plank's constant). Sharvin[2] realized that this resistance should appear at the interface between the device and its contacts, where the number of conduction modes changes sharply. This Landauer-Sharvin resistance thus sets the ultimate resistance limit for ballistic electrons and gives a performance bound on real-life devices, where electrons are forced to transition frequently between metals and semiconductors.

A growing body of theoretical[21–32] and experimental[3–18] evidence suggests that when the interaction between electrons is sufficiently strong to dominate their scattering, the electronic system behaves as a hydrodynamic fluid. Key hydrodynamic features have been observed in transport[3–8,10,11,15,16] and imaging[9,12–14,17,18] experiments. Interestingly, transport measurements of hydrodynamic electrons flowing through constrictions[7] observed that they conducted up to 15% better than their ballistic counterparts. Theoretically, this was explained[28] by hydrodynamic lubrication of the constriction walls. This raises a question of fundamental and practical importance[19,20]: if the Landauer-Sharvin resistance limits ballistic electrons, what is the ultimate conduction limit for hydrodynamic electrons?

In this work, we show experimentally that hydrodynamic electrons can dramatically outperform ballistic electrons' limitations. By imaging electronic flows in a Corbino disk geometry, we observe that ballistic electrons exhibit roughly half of their Landauer-Sharvin resistance being distributed through the bulk of the device rather than at the contacts' interfaces. At elevated temperatures, we find that electron hydrodynamics efficiently eliminates this 'bulk Landauer-Sharvin' resistance. This observation is consistent with the recent theoretical prediction[20] that hydrodynamic electrons can flow without Landauer-Sharvin resistance. By adding a small magnetic field, we set the electrons into a spiraling motion, generating a viscous boundary layer near the contacts.

This layer provides the first real-space observation of the key emergent length scale of hydrodynamic theories - the Gurzhi length. Our findings demonstrate that hydrodynamics can dramatically modify the well-established rules for electrons obeyed by their ballistic counterparts.

To understand the origin of the Landauer-Sharvin resistance, consider ballistic transport through two device geometries: a straight channel (Fig. 1a) and a Corbino disk (Fig. 1b). In both cases, the total device resistance is given by the Landauer-Sharvin resistance, determined by the number of conduction modes traversing the device. In a straight channel, this resistance is given by $R_{sh} = \frac{\pi h}{4e^2}\frac{1}{k_F W}$, where $k_F$ is the Fermi momentum and W is the channel width. In a Corbino geometry, the width is replaced by the circumference of the inner contact, $2\pi r_{in}$, and its total resistance is $R_{sh}^{in} = \frac{\pi h}{4e^2}\frac{1}{k_F(2\pi r_{in})}$. In a straight channel, this resistance drops only at the contact interfaces, with half dropping at each contact interface and none in the bulk (Fig. 1c). The situation changes in a Corbino geometry (Fig. 1d): here, similar to a straight channel, half of the Landauer-Sharvin resistance drops at the inner contact interface. However, curiously, theory predicts[19,20] (Fig. 1d) that the other half should be distributed across the bulk of the device.

The theoretical prediction that the Landauer-Sharvin resistance can be distributed across the bulk of a device highlights it's geometrical/phase-space origin – this resistance appears whenever there is a spatial gradient in the number of conduction modes. In a straight channel, the number of modes changes sharply at the contact interfaces but is fixed throughout the bulk (Fig. 1e). Consequently, the Landauer-Sharvin resistance appears only at the contact interfaces. In contrast, in a Corbino disk, the number of conduction modes gradually decreases with decreasing radius (Fig. 1f). An electron traveling from the outer to the inner contact thus experiences a gradually shrinking phase space, which should manifest as a distributed bulk Landauer-Sharvin resistance.

Interestingly, recent theory[19] has suggested that for hydrodynamic electrons in a Corbino geometry, the bulk Landauer-Sharvin resistance, which forms about half of the total device resistance, should vanish. A new theory[20] analyzed generalized flow geometries and showed that the Landauer-Sharvin resistance appears whenever there is a gradient in the number of conducting modes, and that it originates from the reflection enforced on electrons whose mode is terminated. Electron-electron scattering allows these electrons to smoothly transfer from a mode that is about to be terminated to a propagating mode, and thus reduces the resistance. Consequently, the theory predicted that for hydrodynamic electrons, resistance should occur only where the number of modes has a non-zero second spatial derivative [20]. Since in a Corbino disk the number of modes increases linearly with distance, this geometry is the ideal testbed to examine the elimination of the ballistic Landauer-Sharvin resistance for hydrodynamic flows.

The devices studied here consist of high mobility hBN-encapsulated monolayer graphene patterned into a Corbino disk geometry, with a graphite back gate that tunes the carrier density, $n$. The graphene spans a disk between radii $r_{in}$ and $r_{out}$, where it connects to inner and outer Cr/Au contacts. The line connecting to the inner contact is deposited over the top hBN layer and a patch of crossed-linked resist, so the graphene disk is not perturbed, preserving its full angular symmetry (SI section 1). In the main text, we present data from a device with $r_{in} = 2\ \mu m$ and $r_{out} = 9\ \mu m$ (optical image in Fig. 1g). Similar results were obtained on a second Corbino device with different dimensions (SI section 8).

A major advantage of the Corbino device geometry over the more commonly used Hall-bar devices is the absence of etched edges and lithographically-defined voltage probes. This eliminates spurious scattering at lithographic features and edges, allowing to measure the unperturbed electron flow. This advantage comes at a price: transport experiments can only measure the overall 2-probe resistance of the device, and thus cannot decipher how this resistance is distributed in space. To solve this, we use a nanotube-based scanning single electron transistor[33] (SET) to spatially map the potential drop associated with the electronic current flow[34]. We drive an AC current, $I$, between the inner and outer Corbino contacts and use the SET to image the local electrostatic potential modulations at this AC frequency (Fig. 1h). This distinguishes the potential drop associated with the current flow from the static disorder potential, which we measure independently in DC in the absence of current. The spatial resolution of the measurement is limited by the scanning height of the SET above the device ($\sim 800\ nm$). In the figures below, we plot the imaged potential normalized by the total current, $R(x, y) = \phi(x, y)/I$, and define the zero of the potential at the outer contact, $\phi(r_{out}) = 0$. The quantity $R(x,y)$ therefore represents the resistance between the point $(x,y)$ and the outer contact.

Fig. 2a shows a typical measured map of $R(x,y)$. Visibly, $R(x,y)$ rises monotonically from the outer to the inner contact. The rise is steeper at the graphene-contact interfaces ($r = r_{in}$ and $r = r_{out}$). Plotted as a colormap (inset), we see that $R(x,y)$ exhibits excellent angular symmetry, attesting to the high quality of the measured device. A similar level of angular symmetry is observed for the different temperatures, densities, and magnetic fields used throughout this work (SI section 3). This symmetry allows us to average over the angular direction and obtain an accurate radial resistance profile, $R(r)$, which conveys the essential information about the nature of the electron flow.

We begin with measurements of the resistance profile at low temperatures, where the transport is expected to be ballistic. Fig. 2b shows $R(r)$ measured at $T = 6\ K$ and $n = 4.5 \times 10^{11}\ cm^{-2}$ (similar phenomenology is also observed at other densities, SI section 7). We see that $R(r)$ starts flat at the outer contact, rises rapidly around $r = r_{out}$, climbs gradually throughout the bulk of the Corbino disk, then rises rapidly again around $r = r_{in}$, and finally flattens out at the inner contact. The overall resistance (the 2-probe resistance that would be measured in transport) can be read out directly from this graph to be $R_{tot} = 19.5\ \Omega$. To compare, the resistance of a completely ballistic Corbino device with perfect contacts is expected to be the Landauer-Sharvin resistance, $R_{sh}^{in}$. Plotting the same measurement but now normalized by $R_{sh}^{in}$ (blue curve, Fig. 2c), we see that our device is not far from this ideal limit, $R_{tot}/R_{sh}^{in} = 1.42$.

The spatially resolved measurement now allows us to break the total resistance into the constituent contact and bulk components. The resistance of the graphene-contact interfaces leads to step functions of $R(r)$ at $r = r_{in}$ and $r = r_{out}$, but these are slightly smeared due to the finite resolution of our imaging. Using an complementary measurement on the same device, we accurately determine the point-spread-function (PSF) of our imaging (SI section 4). Thus, the only free parameter in these contact step functions, shown by $R_c^{in}(r)$ and $R_c^{out}(r)$ in Fig. 2c, is their height.

Remarkably, if we fit the measured $R(r)$ (blue, Fig. 2c) to $R_{LS}(r) + R_c^{in}(r) + R_c^{out}(r)$ (dotted black, Fig. 2c), where $R_{LS}(r) = \frac{R_{sh}^{in}}{\pi}\mathrm{asin}\left(\frac{r_{in}}{r}\right)$ is the theoretically-predicted dependence of the Landauer-Sharvin bulk resistance[19,20], we find extremely close agreement throughout the bulk (~10 % difference). Note that the smeared contact functions, $R_c^{in}(r)$ and $R_c^{out}(r)$, penetrate very little into the bulk and therefore do not affect the quality of the fit in the bulk. The close agreement that we see is especially impressive given the fact that the expression for the Landauer-Sharvin bulk resistance has no free parameters. This measurement therefore provides the first real-space evidence of a distributed bulk Landauer-Sharvin resistance, originating from a spatial gradient of the number of conduction modes.

The small difference between the measurement and the ideal Landauer-Sharvin expression can be readily accounted for by weak impurity scattering. Such scattering leads to an ohmic term that depends logarithmically on $r$, $R_{ohm}(r)/R_{sh}^{in} = \frac{2r_{in}}{\pi l_{MR}} \log\left(\frac{r}{r_{in}}\right)$, with one free parameter – the momentum-relaxing mean free path, $l_{MR}$. Adding this to $R_{LS}(r)$, we obtain an excellent fit to the measurement with $l_{MR} = 40 \ \mu m$ (dashed red, Fig. 2c). Such a long mean free path is consistent with previous measurements on high mobility graphene devices[14,35], and is much longer than the Corbino channel length ($r_{out} - r_{in} = 7\mu m$), explaining the smallness of the ohmic contribution to the total resistance.

From the above fit we also obtain the magnitude of the graphene-contact interface resistances. The obtained inner contact resistance step height is $0.82R_{sh}^{in}$, larger than the $0.5R_{sh}^{in}$ expected for an ideal contact (Fig. 1d). This difference reflects a contact transparency of $T_{in} \sim 0.75$ (SI section 5), on par with the best transparencies achieved with graphene contacts[35,36]. By subtracting the fitted contact resistance steps from the measured profile, we obtain $R_{bulk}(r) = R(r) - \left(R_c^{in}(r) + R_c^{out}(r)\right)$ (inset, Fig. 2c). This quantity gives the most accurate description of the bulk resistance profile, even very close to the contacts, because it eliminates the smeared tails of the contacts step functions. In the remainder of the paper, we will use this quantity to explore the physics in the bulk.

The measured dependence of $R_{bulk}(r)$ on carrier density is shown in the left inset of Fig. 2d. We see that the total bulk resistance varies strongly with density (~ factor 4 over the measured density range). However, recalling that the number of conduction channels scales as $k_F \sim \sqrt{n}$, and normalizing each curve by $R_{sh}^{in}$ at the corresponding density, we find that all curves collapse to a similar dependence (Fig. 2d, main panel) close to the Landauer-Sharvin expression (dotted). This demonstrates that at $T = 6\ K$ the Landauer-Sharvin bulk resistance is the dominant contribution over a wide range of carrier densities. With decreasing $n$, the curves depart further from the ballistic limit, pointing to a growing ohmic component. Using a fit to $R_{LS}(r) + R_{ohm}(r)$, we obtain the $n$-dependence of $l_{MR}$ (Fig. 2d right inset), in good agreement with previous measurements[12,14,35].

Having established the behavior at low temperatures, we proceed to explore the flow at elevated temperatures. Increased temperature increases both electron-phonon (e-ph) and electron-electron scattering. The former is momentum-relaxing and is thus expected to increase the device's resistance. Since the geometric and ohmic contributions are additive (as shown below), one may expect the total resistance to increase with the added ohmic resistance. For example, at $T \sim 140\ K$, the previously measured[14,35,37] e-ph mean free path ($\sim 4\ \mu m$) implies that the total resistance should more than double.

The measured temperature dependence in Fig. 3a shows a surprisingly different behavior. The figure plots the measured $R_{bulk}(r)$ at temperatures $T = 6\ K$ to $140\ K$, where in all the curves we subtracted the same contact resistance steps, those obtained from the fit at $T = 6\ K$ (gray traces, Fig 2c). We see that instead of increasing as $T$ increases, the resistance first decreases up to $T \approx 60\ K$ and then only mildly increases ($\sim 20\%$) toward $T = 140\ K$ (Fig. 3a inset). The measured spatial dependence of $R_{bulk}(r)$, plotted on a logarithmic $r$ axis in Fig. 3a, provides a hint for the underlying physics: whereas at low $T$, the dependence is curved, as expected from an $\mathrm{asin}(r_{in}/r)$ dependence (bottom dashed line), at $T = 140\ K$ the dependence follows a perfectly straight line throughout almost the entire bulk of the device (top dashed line). Namely, at elevated $T$ the resistance follows a pure $\log(r/r_{in})$ dependence. This suggests that as the ohmic e-ph contribution builds up, the contribution of the Landauer-Sharvin bulk resistance disappears, explaining why the resistance doesn't double as naively expected.

Given the unavoidable presence of e-ph scattering at elevated temperatures, even in the cleanest graphene samples, how can we resolve the clean-limit hydrodynamic flow profiles, namely those involving only electron-electron and not momentum-relaxing collisions? Here, the angular symmetry of the Corbino geometry proves advantageous. In the presence of this symmetry and within the relaxation time approximation there is a direct mapping between the flow with ohmic scattering and the clean-limit flow: If $R_{bulk}(r)^{L_{MR}}_{L_{ee}}$ is the profile with momentum-conserving and momentum-relaxing mean free paths $L_{ee}$ and $L_{MR}$, then one can obtain from it the clean-limit profile, $\tilde{R}_{bulk}(r)$, by a mere subtraction of an ohmic term (the proof is given in SI section 9):

$$\tilde{R}_{bulk}(r)_{L_{ee}=(l_{ee}^{-1}+l_{MR}^{-1})^{-1}}^{L_{MR}=\infty} = R_{bulk}(r)_{L_{ee}=l_{ee}}^{L_{MR}=l_{MR}} - \frac{\hbar}{2e^2 k_F l_{MR}}\log(r/r_{in}) \quad (1)$$

The last term has only one free parameter, $l_{MR}$, which we obtain directly from independent magneto-hydrodynamic imaging experiments. Before discussing these experiments, we first substitute the obtained $l_{MR}$ into equation (1), giving us directly $\tilde{R}_{bulk}(r)$ without adding any free parameters.

Figure 3b plots the obtained clean-limit hydrodynamic flow profiles, $\tilde{R}_{bulk}(r)$, at the various temperatures. At the lowest temperature, $\tilde{R}_{bulk}(r)$ follows the bulk Landauer-Sharvin dependence. With increasing $T$, however, this geometrical resistance gradually disappears. Remarkably, at $T = 140\ K$ the profile becomes completely flat throughout most of the bulk of the device, apart from small regions ($< 1\ \mu m$) near the inner and outer contacts.

To understand these measurements in more detail, we performed numerical Boltzmann calculations for a Corbino geometry with an electron–electron mean free path $l_{ee}$, taken within the relaxation time approximation (SI section 10). Fig. 3c shows the calculated $\tilde{R}_{bulk}(r)/R_{sh}^{in}$ for various $l_{ee}$ values, smeared with the experimental PSF. For $l_{ee} = \infty$ the calculation recovers the Landauer-Sharvin dependence. With decreasing $l_{ee}$ the Landauer-Sharvin resistance is gradually reduced. Once the mean free path becomes much shorter than the channel length, $l_{ee} \ll r_{out} - r_{in}$, the resistive drop occurs only at a distance $\sim l_{ee}$ from the contacts, nicely matching the measurements at elevated temperatures. This is the hydrodynamic buildup distance, over which the electron-electron interactions rearrange the flow from ballistic to hydrodynamic, and its accumulated resistance is the Stokes resistance[19] $\sim l_{ee}/k_F r_{in}^2$. The calculations also reproduce the appearance of a resistive outer contact step with increasing $T$, albeit stronger than observed experimentally (SI section 11). Most importantly, similar to the experimental result, we see that throughout most of the bulk of the disk, the Landauer-Sharvin geometrical resistance is completely eliminated by the hydrodynamic flow.

Finally, we turn to exploring hydrodynamic flow at non-zero magnetic fields. Corbino geometry is an ideal testbed for magneto-hydrodynamics due to the lack of disruptive physical edges. Fig. 4a shows the evolution of $R_{bulk}(r)$, measured at $T = 140\ K$, with a perpendicular magnetic field, $B$, ranging from zero to $30\ mT$. Evidently, the magnetic field increases the resistance throughout the Corbino channel. Plotting $R_{bulk}$ vs. $B$ at several radii (inset), we see positive $\sim B^2$ magnetoresistance (dashed lines).

Figure 4b presents the ratio between the radial electric field, obtained via a numerical derivative of the measured potential, $E_r = \frac{d\phi}{dr}$, and the radial current density, $j_r = I/2\pi r$. In contrast to a Hall bar geometry, where such a ratio between longitudinal field and current density gives the longitudinal resistivity, $\rho_{xx}$, in a Corbino geometry, this ratio yields the inverse longitudinal conductivity, $\sigma_{xx}^{-1}$. This is because, in the latter, due to angular symmetry the transverse field rather than the transverse Hall current is zero. We see that at $B = 0$, $\sigma_{xx}^{-1}$ is independent of $r$ throughout most of the disk's bulk. With increasing $B$ the magnitude of $\sigma_{xx}^{-1}$ increases. Interestingly, this increase is not constant in space, but is larger at the center of the conducting channel and smaller at its sides (dashed arrows).

Recalling that $\sigma_{xx}^{-1} = \rho_{xx} + \rho_{xy}^2/\rho_{xx}$ and $\rho_{xy} \sim B$, we see that the inverse conductivity, while being equal to the resistivity at $B = 0$, acquires an additional Hall component at finite $B$. This term arises from the appearance of an angular current density, $j_\theta$, and a corresponding Hall angle, $\tan(\theta_{Hall}) = \rho_{xy}/\rho_{xx} = j_\theta/j_r$ (Fig. 4c inset), both of which grow linearly with $B$. By fitting $\sigma_{xx}^{-1}(r, B)$ to $\rho_{xx}(r) + a(r)B^2$ we get directly from $a(r)$ the dependence of $\tan(\theta_{Hall})/B$ on $r$, plotted in Fig. 4c. Visibly, $\tan(\theta_{Hall})$ is maximal at the center of the channel, but drops gradually toward the contacts. The length scale of the drop is much longer than our resolution limit (black line, Fig. 4c). Measurement of $\tan(\theta_{Hall})$ in the full 2D plane (Fig. 4d), shows a similar behavior: the Hall angle is largest at the center of the conducting channel, and gradually drops toward the contacts. Thus, in a finite magnetic field we observe a length scale that does not exists at $B = 0$, and describes the spatial change of $\theta_{Hall}$.

Such an emergent scale was proposed by recent theories of magneto-hydrodynamic flow in a Corbino geometry[19,31], whose predicted flow lines are reproduced in Fig. 4e.

These lines are slanted with respect to the radial electric field by the local Hall angle, $\tan(\theta_{\text{Hall}}(r)) = \frac{j_\theta}{j_r}$. At the center of the conducting channel, the Lorentz force and electric field balance to give the standard expression for the Hall angle, $\tan(\theta_{\text{Hall}}) = \frac{l_{MR}}{R_c}$ ($R_c$ is the cyclotron radius). However, near the contacts, $\theta_{\text{Hall}}$ must go to zero, since the electrons are injected from the contact isotropically, namely, $j_\theta = 0$. Formally, the theory gives:

$$\tan(\theta_H) = \frac{l_{MR}}{R_C} + C_1 r I_1\left(\frac{r}{l_G}\right) + C_2 r K_1\left(\frac{r}{l_G}\right), \tag{2}$$

where $I_1, K_1$ are the modified Bessel functions, $C_1, C_2$ are constants chosen such that $\tan(\theta_H) = 0$ at $r = r_{in}, r_{out}$, and $l_G = \sqrt{l_{MR} l_{ee}/4}$ is the Gurzhi length. The red curve in Fig. 4c plots the profile for $l_{MR} = 4.35\ \mu m$ and $l_{ee} = 1.3\ \mu m$, which agrees best with the experiment. In clean samples and elevated temperatures, $l_{MR}$ is dominated by electron-phonon coupling and is thus rather universal. Indeed the $l_{MR}$ we find here is in full agreement with earlier experiments[12,14,35]. Moreover, the $l_{ee}$ obtained from the $B = 0$ experiment (Fig. 3) and from the magneto-hydrodynamic experiment (Fig. 4) closely agree, although they have completely different manifestations in these two flow regimes: in the former, $l_{ee}$ gives the hydrodynamic buildup length and the Stokes resistance of the inner contact (Figs. 3b and 3c). In the latter, $l_{ee}$ enters only through its geometrical average with $l_{MR}$ to give the spatial scale of the $\theta_H$ gradient. Finally, we note that this is the first time that this key emergent length of electron hydrodynamics theory, the Gurzhi length, has been observed directly.

Our experiments demonstrate the intimate connection between the Landauer-Sharvin resistance and the spatial gradient in the number of conduction modes. Corbino devices have such gradients naturally, by virtue of their geometry. It is interesting to note that these devices are mathematically equivalent to devices with a simple (e.g., rectangular) geometry, in which the spatial gradient in the number of modes is caused by chemical doping rather than geometry[20]. This means that the physics discussed in this paper might even be relevant for future real-world devices. The observation that hydrodynamic electrons can dramatically outstrip the fundamental bounds of their ballistic counterparts, is thus of fundamental as well as technological importance.

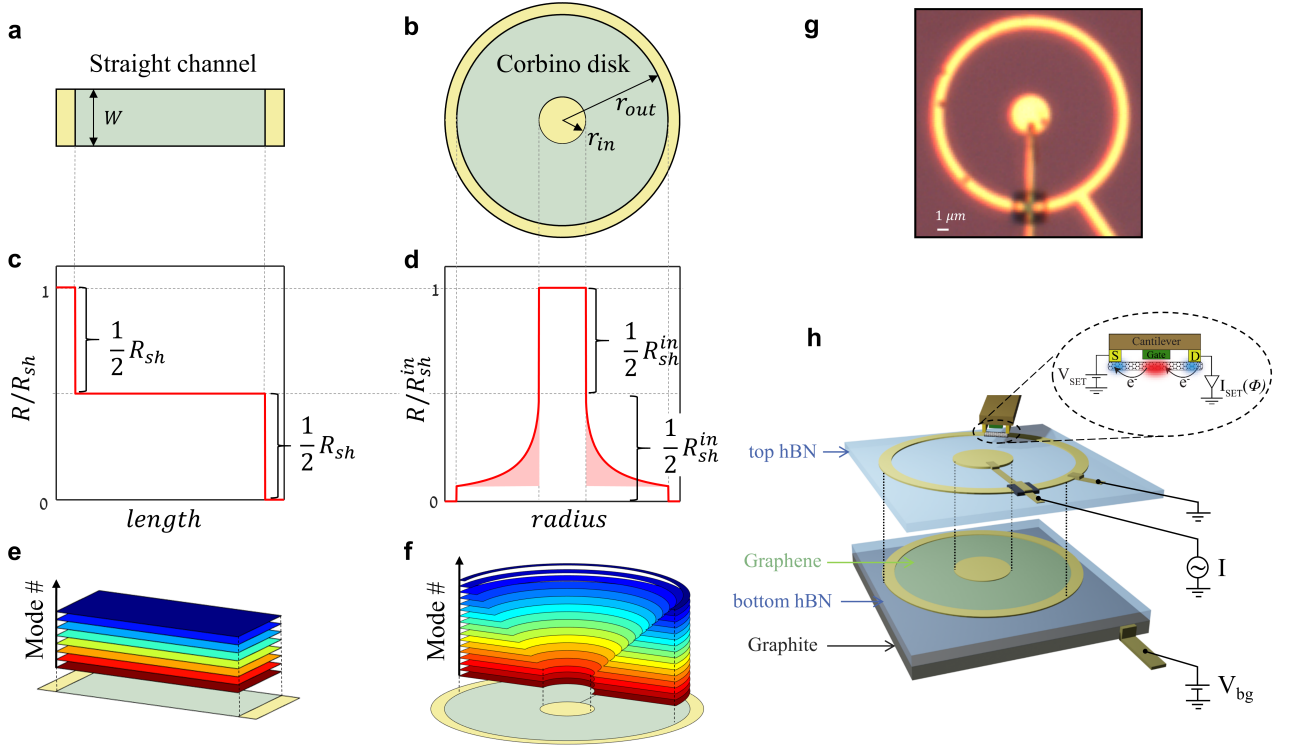

**Figure 1: Landauer-Sharvin bulk geometrical resistance and the experimental setup for its measurement.** Comparing two channel geometries: **a**. A straight channel (channel - green, contacts - yellow) **b**. Corbino disk channel (inner and outer radii are $r_{in}$ and $r_{out}$, respectively). In the ballistic limit (no scattering, perfect edges) the total (2-probe) resistance of both geometries is inversely proportional to the number of conduction modes traversing the device, multiplied by $e^2/h$. For a straight graphene channel this gives $R_{sh} = \frac{\pi h}{4e^2} \frac{1}{k_F W}$ ($W$ is the channel width and $k_F$ is the Fermi momentum). For a Corbino disk, $W$ is replaced by the inner contact circumference, giving a total resistance of $R_{sh}^{in} = \frac{\pi h}{4e^2} \frac{1}{k_F (2\pi r_{in})}$. **c.** In the straight channel, half of the Landauer-Sharvin resistance drops sharply at each contact interface and there is no resistive drop in the bulk. **d**. In the Corbino disk, similarly, half the Landauer-Sharvin resistance drops on the inner contact. However, the other half is distributed throughout the bulk of the Corbino disk, falling off as $R(r) = \frac{1}{2} R_{sh}^{in} \frac{2}{\pi} \mathrm{asin}(r_{in}/r)$. The Landauer-Sharvin resistance has a fundamental geometrical / phase-space origin – it appears whenever there is a spatial gradient in the number of conduction modes: **e**. In a straight channel the number of modes is constant throughout the bulk but changes sharply at the interfaces with the metallic contacts that effectively have an infinite number of modes. **f**. In a Corbino disk, there is a similar sharp change in the mode number at the inner contact interface, but throughout the bulk there is a linear increase of mode number with the radius, leading to the bulk Landauer-Sharvin resistance. **g**. Optical image of one of the studied devices ($r_{in} = 2\ \mu m$, $r_{out} = 9\ \mu m$). **h**. The device is composed of top hBN, graphene, bottom hBN and a graphite back gate, with inner (circular) and outer (ring) contacts. The carrier density is tuned by voltage $V_{bg}$ on the graphite back-gate. We use a nanotube-based single electron transistor (SET) (inset) to image the potential in the device while flowing a current $I$ between the contacts.

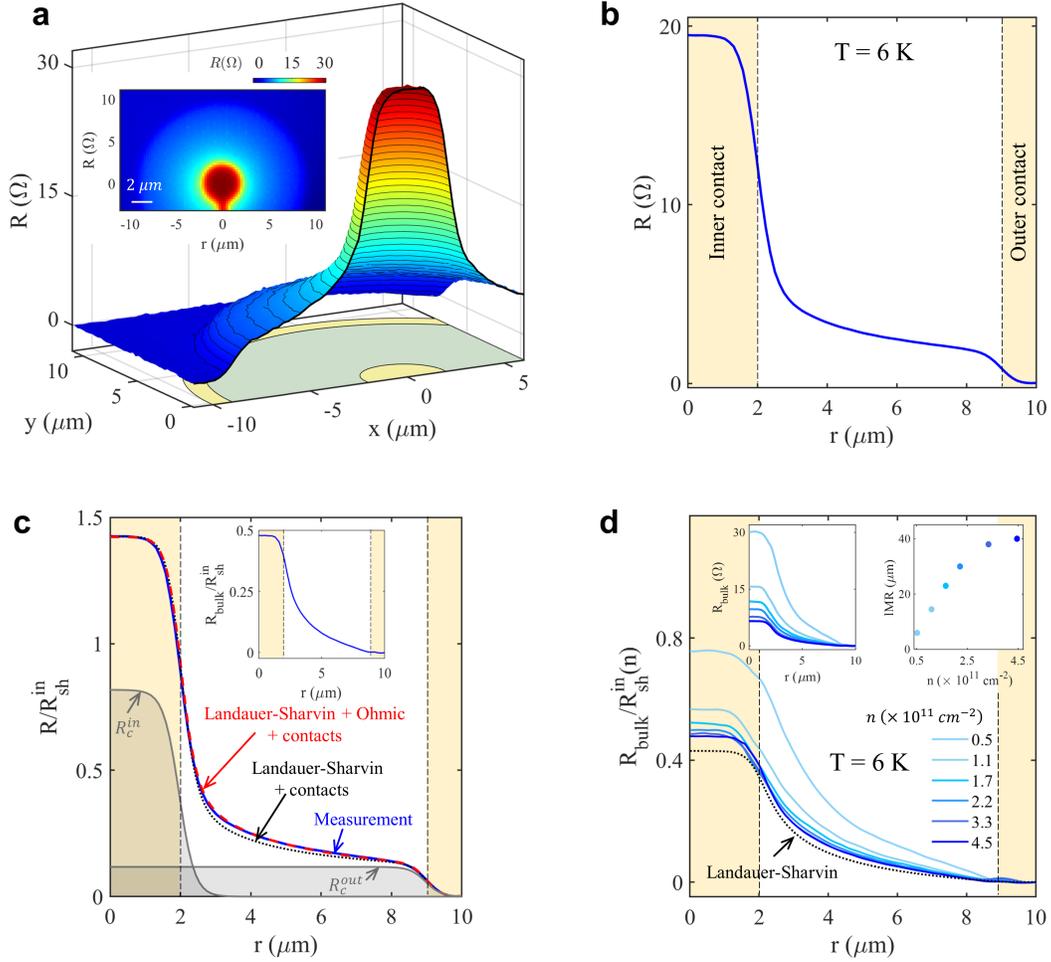

**Figure 2: Imaging the Landauer-Sharvin bulk resistance in a Corbino disk**. **a**. Spatially resolved image of the resistance $R(x,y) = \phi(x,y)/I$, where $\phi(x,y)$ and $I$ are the measured potential and current, respectively, displayed above the schematic of the Corbino device. ($T = 140\,K$, $n = 4.5 \times 10^{11}\,cm^{-2}$). Inset: the same measurement presented as a colormap. The measurement exhibits excellent angular symmetry, allowing us to average along the angular direction and obtain $R(r)$. **b.** The measured radial dependence of the resistance, $R(r)$, at $T = 6\,K$, $n = 4.5 \times 10^{11}\,cm^{-2}$. Contacts are marked yellow and their interface with the graphene by dashed lines. **c**. Disentangling the different components of the resistance. The blue curve is the same measurement as in panel b, but now plotted normalized by $R_{sh}^{in}$. We fit this curve with a function that include bulk and contact dependence. The graphene-contact interface resistances are described by $R_c^{in}(r)$ and $R_c^{out}(r)$ (gray curves), which are step functions at $r = r_{in}$ and $r = r_{out}$, smeared by the point-spread-function (PSF) of our imaging experiment, which is measured separately in an complementary experiment (SI section 4). The dashed dotted line shows a fit of the measurement to $R_{LS}(r) + R_c^{in}(r) + R_c^{out}(r)$, where $R_{LS}(r) = \frac{R_{sh}^{in}}{\pi} \mathrm{asin}\left(\frac{r_{in}}{r}\right)$ is the theoretically-predicted Landauer-Sharvin bulk geometrical resistance that has no free parameters. The dashed red line is a fit to a similar function which includes in addition an ohmic term, $R_{ohm}(r)/R_{sh}^{in} = \frac{2r_{in}}{\pi l_{MR}} \log\left(\frac{r}{r_{in}}\right)$, with a momentum-relaxing mean free path of $l_{MR} = 40\,\mu m$. Inset: The bulk component of the resistance, obtained from the measured $R(r)$ by subtracting away the fitted contact resistance curves, $R_{bulk}(r) = R(r) - \left(R_c^{in}(r) + R_c^{out}(r)\right)$. **d**. Left inset: Measured $R_{bulk}(r)$ at various carrier densities, $n$ (see key). Main panel: Same curves, but normalizing each curve by the Sharvin resistance at the corresponding density, $R_{sh}^{in}(n)$. Dotted line, theoretical Landauer-Sharvin bulk dependence. Right inset: $l_{MR}$ vs. $n$ obtained from fitting the graphs in the main panel to $R_{LS}(r) + R_{ohm}(r)$.

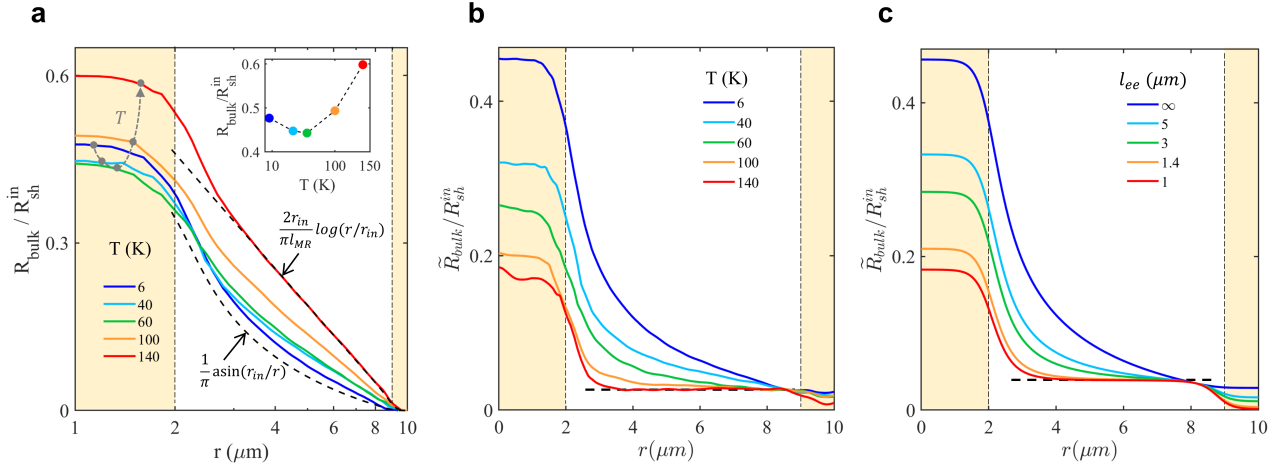

**Figure 3. Observation of the perfect elimination of Landauer-Sharvin bulk resistance by hydrodynamic electron flow**. **a**. Measured $R_{bulk}(r)$ at various temperatures, $T$, (see key) normalized by $R_{sh}^{in}$ and plotted on a logarithmic $r$ axis. Similar to Fig 2, $R_{bulk}(r)$ is obtained from the measured $R(r)$ by removing the contact resistance contribution. Note that we removed from the curves at the different temperatures the same contact resistance traces, those obtained from the fit at $T = 6\ K$ ($R_c^{in}(r)$ and $R_c^{out}(r)$ shown in gray in Fig. 2c). We see that the total bulk resistance first decreases with increasing $T$ and then only mildly increases (gray dashed line). Inset: total bulk resistance normalized by $R_{sh}^{in}$ as a function of $T$. The $r$ dependence of the resistance evolves from Landauer-Sharvin dependence (bottom dashed line), at $T = 6\ K$, to a purely logarithmic dependence (top dashed line), at $T = 140\ K$, with a small deviation only very close to the inner contact ($< 1\ \mu m$). **b**. The clean-limit hydrodynamic flow profiles, $\tilde{R}_{bulk}(r)$, at various $T$'s, obtained from the data in panel a using equation (1) and a momentum-relaxing mean-free path, $l_{MR}$, measured by a completely independent experiment on the same device (see main text). Note that in the bulk of the device these curves involve no free parameters. Notably, the distributed Landauer-Sharvin bulk resistance at $T = 6$ K is fully eliminated throughout most of the disk's bulk once the temperature has reached $T = 140$ K (the horizontal dashed black line is a guide to the eye). A small resistive component remains only close ($< 1\ \mu m$) to the inner contact. In addition, we see a slight increase of outer contact resistance with $T$ **c**. Theoretical clean-limit hydrodynamic profiles, calculated using Boltzmann equations in the Corbino geometry. Different traces correspond to different electron-electron scattering length, $l_{ee}$ (see key). Similar to the subtraction of the $T = 6\ K$ contact resistance steps in the experimental data, here we subtracted from all the curves the contact resistance steps calculated at $l_{ee} = \infty$. To allow for quantitative comparison with the experiment, the theory curves are smeared with the measured PSF of the imaging experiment. We observe a close correspondence between experiments and theory in the detailed profile shapes (see text).

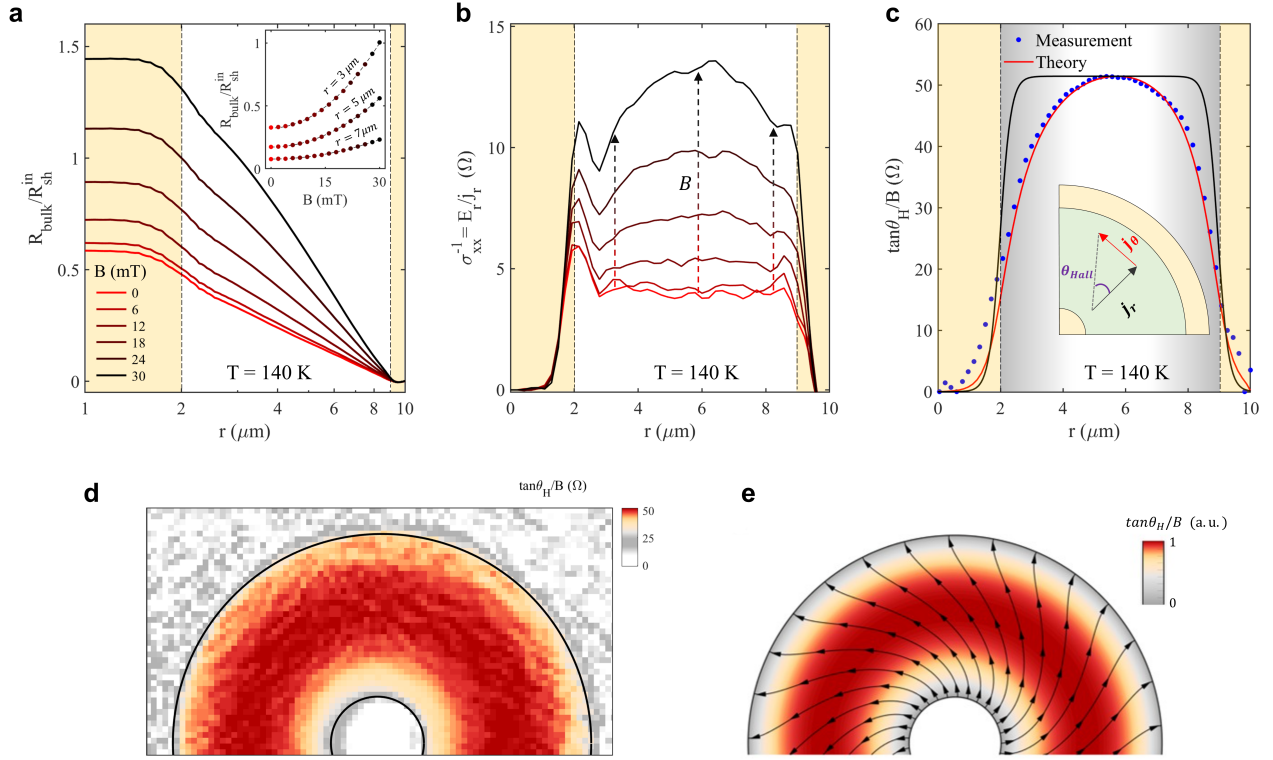

**Figure 4. Imaging spiraling magneto-hydrodynamic electron flow and its Gurzhi boundary layer**. **a.** $R_{bulk}(r)$, normalized by $R_{sh}^{in}$, measured at $T = 140\ K$ and various perpendicular magnetic fields, $B$ (see key). Inset: $R_{bulk}/R_{sh}^{in}$ at three spatial locations, $r = 3, 5, 7\mu m$, measured as a function of $B$. Dashed lines are parabolic fits. **b.** The inverse conductivity, $\sigma_{xx}^{-1} = E_r/j_r$, vs. $r$ at various $B$'s (same key as in panel a). $E_r$ is the radial component of the electric field, obtained by numerically differentiating the measured potential, $E_r = d\phi/dr$, and $j_r = I/2\pi r$ is the current density obtained from the measured total current, $I$. Note that, while $\sigma_{xx}^{-1}$ is independent of $r$ throughout most of the disk's bulk at $B = 0$, the component added to $\sigma_{xx}^{-1}$ at non-zero $B$ is largest in the center of the Corbino channel and decays gradually toward the contacts (dashed arrows). **c.** Spatial dependence of the Hall angle, $\theta_H$, obtained from fitting the quadratic-in-$B$ term from panel b. The figure plots $\tan(\theta_H)/B$ (since $\theta_H$ increases linearly with $B$) as a function of $r$. While $\theta_H$ plateaus at the center of the channel, it drops gradually toward zero at the contacts. The drop happens over the Gurzhi length (gray shading). This length scale is considerably longer than our imaging resolution as is apparent from the black curve, which is a rectangular function convolved with our imaging PSF. The red curve is a fit to hydrodynamic theory (see text. Inset: diagram sketching the radial current density, $j_r$, the angular current density, $j_\theta$, and the Hall angle, $\theta_H$. **d.** $\tan(\theta_H)/B$, but now shown in a full 2D imaged spatial map. To obtain this image we measured $R(x, y)$ maps at $B = 0, 6, 12, 18, 24, 30\ mT$, then for each $(x, y)$ point determined the quadratic-in-$B$ term in the resistance, from which we obtain the local $\tan(\theta_H)/B$. Similar to panel c, which is the angular average of this measurement, even in this spatially resolved map we can clearly see that $\theta_H$ plateaus at the center of the channel, but drops gradually to zero at the contacts, over a rather long spatial scale. **e.** The calculated map of $\tan(\theta_H)/B$ using Navier-Stokes magneto-hydrodynamic equations and the parameters of the experiment in panels c and d. Overlaid are the flow lines. At the center of the channel, the flow lines are skewed from the radial direction following the standard expression for the Hall angle $\tan(\theta_H) = l_{MR}/R_c$. ($R_c$ is the cyclotron radius). The boundary condition at the contacts dictates that $j_\theta = 0$ and thus $\theta_H = 0$. The climb of $\theta_H$ from zero to its bulk value occurs over the Gurzhi length, $l_G = \sqrt{l_{ee}l_{MR}/4}$, corresponding closely to the length scale that emerged in the experiment.

## Methods:

Device fabrication: Scanning SET devices were fabricated using a nanoscale assembly technique[38]. The graphene/hBN devices were fabricated using electron-beam lithography and standard etching and nanofabrication procedures[35] to define the channels and evaporation of Cr/Au (S4) to deposit contact electrodes.

Measurements: The measurements are performed on multiple graphene devices in home-built, variable temperature, Attocube-based scanning probe microscopes. The microscopes operate in vacuum inside liquid helium dewar with superconducting magnets, and are mechanically stabilized using Newport laminar flow isolators. A local resistive SMD heater is used to heat the samples under study from $T = 7.5$ K to $T = 150$ K, and a DT-670-BR bare chip diode thermometer mounted proximally to the samples and on the same printed circuit boards is used for precise temperature control. The voltage imaging technique employed is presented in reference[34]. Voltages and currents (for both the SET and sample under study) are sourced using a home-built DAC array, and measured using a home-built, software-based audio-frequency lock-in amplifier consisting of 1uV accurate DC+AC sources and a Femto DPLCA-200 current amplifier and NI-9239 ADC. The local gate voltage of the SET is dynamically adjusted via custom feedback electronics employing a least squares regression algorithm to prevent disruption of the SET's working point during scanning and ensure reliable measurements.

The voltage excitations applied to the graphene channels were as follows: 1 mV at $T = 6$ K, and 8 mV at $T = 140$ K. The magnetic fields applied ranged between $\pm 30$ mT.

**Acknowledgements:** We thank L. Ella, G. Falkovich, L. Levitov, M. Polini, M. Shavit, A. Rozen, A. V. Shytov and U. Zondiner for useful discussions. Work was supported by the Leona M. and Harry B. Helmsley Charitable Trust grant, ISF grant (1182/21), Minerva grant (713237). TS acknowledges the support of the Natural Sciences and Engineering Research Council of Canada (NSERC), in particular the Discovery Grant [RGPIN-2020-05842], the Accelerator Supplement [RGPAS-2020-00060], and the Discovery Launch Supplement [DGECR-2020-00222]. During the preparation of this manuscript, we became aware of a partially related STM work[17], which images voltage drops in flows across a constriction.

**Contributions:** C.K., J.B., J.A.S, A.K.G. and S.I. designed the experiment. C.K., J.B., J.A.S. performed the experiments. J.B., D.P. fabricated the devices. C.K., J.B., J.A.S, and S.I. analyzed the data. T.S. A.S. and S.I. wrote the theoretical model. K.W. and T.T. supplied the hBN crystals. C.K., J.B., A.S. and S.I. wrote the manuscript with input from other authors.

**Data availability:** The data that support the plots and other analysis in this work are available from the corresponding author upon request.

Supplementary materials for:

# Imaging Hydrodynamic Electrons Flowing Without Landauer-Sharvin Resistance

Chandan Kumar[†], John Birkbeck[†], Joseph A. Sulpizio[†], David J. Perello, Takashi Taniguchi, Kenji Watanabe, Oren Reuven, Thomas Scaffidi, Ady Stern, Andre K. Geim, and Shahal Ilani[*]

## Contents

# S1. Device fabrication

Our devices consist of monolayer graphene encapsulated between two hexagonal boron nitride (hBN) layers, with a graphite flake underneath acting as a back-gate. These devices are assembled using the standard dry transfer technique[1]. Briefly, a polypropylene carbonate (PPC) coated polydimethylsiloxane (PDMS) was used to pick up the top hBN, monolayer graphene, and the bottom hBN, and this stack was then dropped on a graphite flake. After assembly, the heterostructure was annealed under ultra-high vacuum at 500 °C for $\sim 8$ hrs, followed by ironing with an AFM tip in contact mode. Both steps helped clean any residues from the surface of the top hBN and improve the sample's homogeneity by removing bubbles and ripples at the hBN-graphene interface[2]. Fig. S1a shows the optical image of the heterostructure. The monolayer graphene area (marked with a white dashed line) is free of bubbles/wrinkles. Next, we defined the Corbino disk using a standard electron beam lithography process, followed by reactive ion etching (RIE) with $CHF_3$ + $O_2$ and metal deposition of Cr (2 nm) / Au (80 nm) (Fig. S1b).

For making contact to the inner circular disk without electrically shorting to the outer metal ring, a bridge was defined on a small section of outer contact (highlighted by the dotted black line in Fig. S1c). The bridge was made by cross-linking polymethyl methacrylate (PMMA), using a high dose (x1000 regular dose) during electron beam lithography. Finally, in another electron beam lithography step, we defined a lead that passes over the cross-linked PMMA, ensuring that the inner circular disk and the outer contact ring are electrically isolated.

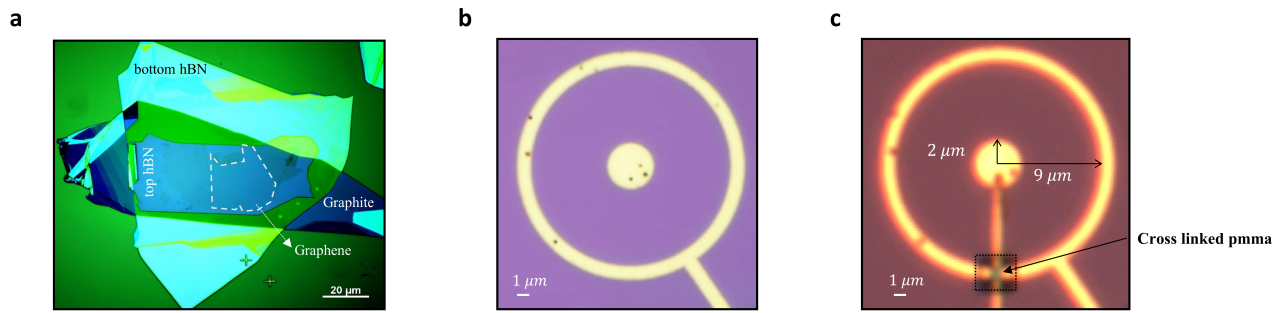

**Fig. S1 | Device fabrication**. **a.** Optical image of the heterostructure (hBN/Graphene/hBN/Graphite) after annealing and AFM ironing. The graphene region is demarcated with a white dashed line and is free of bubbles and wrinkles. The bottom graphite acts as a back-gate. The thicknesses of the top and bottom hBN are $40\ nm$ and $76\ nm$, respectively. **b.** Optical image of the heterostructure after making 1D contacts that define the Corbino disk geometry. **c.** Optical image of the Corbino disk after contacting the inner circular disk. The contact to the inner circular disk passes over cross-linked PMMA, defined over a small segment of outer contact (dashed black), electrically isolating the inner contact from the outer ring. The inner and outer radius of the graphene disk in this device are $r_{in} = 2\ \mu m$ and $r_{out} = 9\ \mu m$ (arrows), respectively.

## S2. Transport measurements

We use standard lock-in techniques to measure the two-probe resistance between the inner and outer contacts of the Corbino disks as a function of carrier density, $n$, and temperature, $T$ (from $6\,K$ to $140\,K$). We find that a significant component of the resistance comes from the lithographic lines that lead to the device. Using the Scanning SET imaging we accurately determine these line resistances by imaging the potential drop between the voltage source and the actual potential of the metal contact, measured by the SET. This line resistance was measured as a function of temperature and back-gate voltage, $V_{BG}$ (which controls the carrier density) and found that it depends on $T$ but not on $V_{BG}$. The measured total line resistance at $T = 6\,K$ is $R_{lines} = 515\,\Omega$ and it increases with increasing temperature, reaching $R_{lines} = 615\,\Omega$ at $T = 140\,K$. We subtract this SET-measured line resistance from the transport-measured two-probe resistance to obtain an effective four-probe resistance. Note that this four-probe resistance still includes the metal-graphene contact resistance, which includes both the fundamental Landauer-Sharvin component, and the component due to imperfect contacts. Fig. S2a plots this effective four-probe resistance as a function of carrier density, $n$, at various temperatures (see legend). Independently, we can determine the total device resistance directly from the SET measurements by reading out from the imaged $R(r)$ (e.g. Fig. 2b, main text) the difference in its value between the outer and inner contacts. The inset to Fig. S2a compares the SET measured device resistance (red dots) and the effective four-probe transport measured device resistance (blue) at $T = 6$ K. We can see an excellent agreement between the two measurements. (Note that the removed line resistance is independent of density and has excellent agreement over the entire density range). Plotting the conductance at $T = 6\,K$ as a function of $\sqrt{n}$ we can estimate the charge inhomogeneity in our samples from the flat region of conductivity near the Dirac point, which comes out to be $\delta n \sim 5 \times 10^9\,cm^{-2}$ (Fig. S2b). In the main text, data is presented for the electron-doped region. Most of the data (apart from Fig. 2d) are taken at a gate voltage of, $V_{BG} = 2\,V$ ($n = 4.5 \times 10^{11}\,cm^{-2}$), though we see similar phenomenology also at other densities (see e.g. section S7 below).

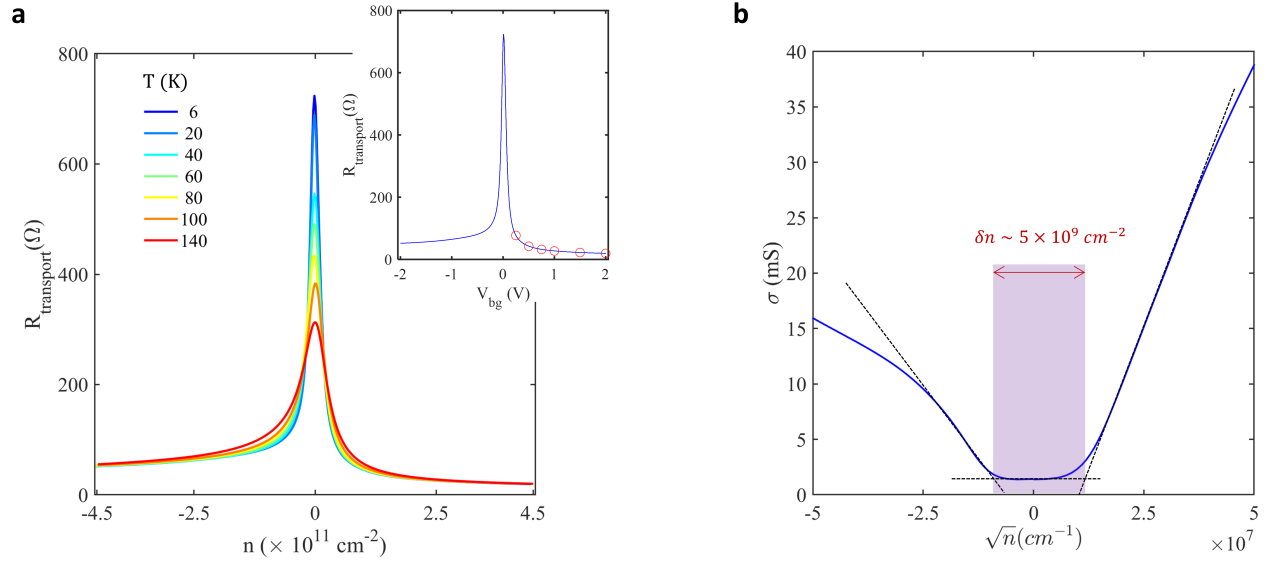

**Fig. S2 | Transport measurements of the Corbino device in the main text**. Main panel: effective four-probe transport resistance, $R_{transport}$, as a function of density, $n$, and at various temperatures, $T$ (see legend). To obtain $R_{transport}$, we use standard lock-in measurements of the two-probe resistance between the Corbino disk's contacts. From this we subtract the lines resistance imaged using the scanning SET. We find that the line resistance depends on temperature but not on the back-gate voltage, $V_{BG}$. Inset: Corbino device resistance as a function of $V_{BG}$ at $T = 6\,K$. The blue line was taken from the main panel (transport measurement with line resistance subtracted) the red dots are obtained from imaging measurements of $R(r)$ using the scanning SET (e.g. Fig. 2b, main text). **b**. Transport measured conductivity, $\sigma$, at $T = 6\,K$, plotted as a function of $\sqrt{n}$. The width of the plateau at the center provides an estimate for the charge disorder in the sample: $\delta n \sim 5 \times 10^9\,cm^{-2}$.

## S3. Angular symmetry of the measured flow

In Figs. 2,3,4 of the main text we plot the resistance profile, $R(r)$, obtained by angular averaging of two-dimensional SET images of $R(x, y)$. This procedure is valid as long as the physics has a high degree of angular symmetry. We demonstrate this symmetry below using a specific measurement and note that a similar level of symmetry exists also for the data measured at other carrier densities and temperatures.

Figs. S3 plots a spatial scan of $R\,(x, y)$ measured at $B = 30\,mT$, $V_{BG} = 2\,V$ and $T = 100\,K$. Similar to the scan in Fig. 2a in the main text, which was performed at a different temperature and magnetic field, also here we see that the measurement exhibits excellent angular symmetry. Panels a and b show the same spatial scan, but with different averaging regions marked by a shaded "pizza" slice (outlined in black and red, respectively, in the two panels). The two resulting $R(r)$ profiles are shown in panel c with the corresponding colors. We can see that the two profiles are practically identical.

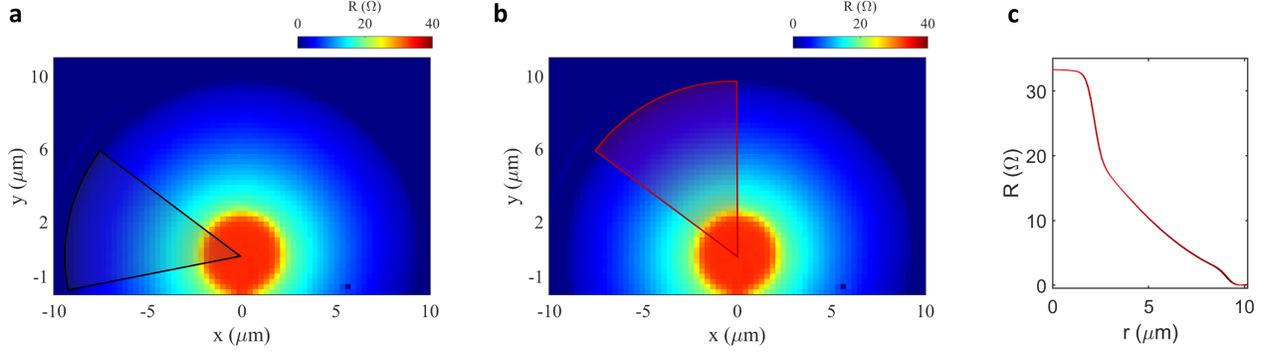

**Fig. S3**: **Angular symmetry of the measured data. a,b** In both panels we show the same spatial map of the resistance, $R(x,y)$, measured at $B = 30\ mT$, $V_{BG} = 2$ V and $T = 100$ K. The shaded slices with black (panel **a**) and red (panel **b**) outline marks the regions used for the angular averaging. **c.** The resistance profiles, $R(r)$, obtained from averaging over the slices in a and b, plotted with corresponding black and red colors.

## S4. Measurement of the point spread function (PSF) of the imaging experiments.

The scanning SET measurements have a finite spatial resolution, determined by the scanning height of the SET above the graphene. This manifests in our scans as a spatial smearing with a point spread function (PSF) that depends on our scanning height. To accurately compare our measurements with the theory we extract the PSF from an independent imaging experiment and convolve the theory curves with this PSF. To determine the PSF we use an experiment that images the workfunction, $W(x,y)$, shown in Fig. S4a, and follow the recipe discussed in our previous work[3]. In contrast to the measurements of $R(x,y)$ which probe the potential that is generated by an electronic current (out of equilibrium), measurement of $W(x,y)$ probe the static (equilibrium) potential, in the absence of current. These are therefore measurements of completely independent properties.

In Fig. S4a we can see that the workfunction is constant throughout most of the graphene disk. On the central gold contact the workfunction is also constant, but with a different value. The transition between these two constants occurs very sharply, over lithographic scales. The workfunction image thus yields a sharp rise, ideal for determining the imaging PSF.

Fig. S4b plots the measured workfunction (blue dots) along the blue dashed radial line in Fig. S4a. Compared to an ideal step function positioned at $r_{in} = 2\ \mu m$ (black line), we see that the measurement is spatially smeared. The red curve shows the step function convolved with a PSF given by $g(r) = 1/\cosh^2\left(1.76\frac{r}{\sigma}\right)$. In our previous work[3] we demonstrated that this PSF describes well the smearing in the experiment and that $\sigma$ corresponds to our scanning height. We find a good fit between the measurement and the step function PSF-smeared with a height value of $\sigma = 0.85\ \mu m$. This is the PSF used in the main text.

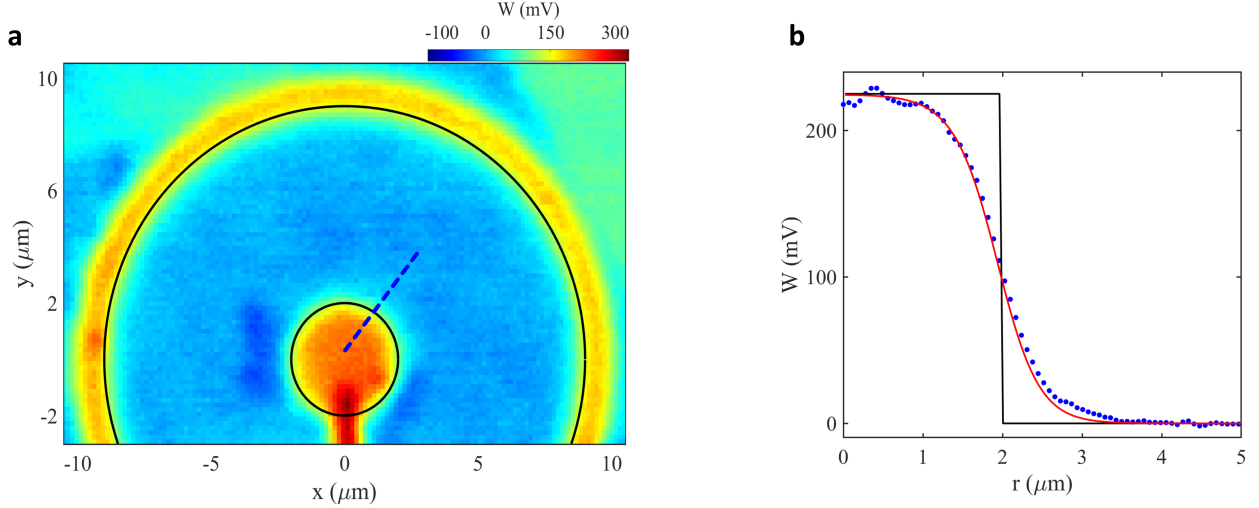

**Fig. S4**: **Obtaining the measurement's PSF from imaging of the workfunction. a.** Colormap of the measured workfunction, $W(x, y)$, over the Corbino disk. The solid black lines indicate the inner and outer radius of the graphene disk. **b.** Measured $W$ (blue dots) along the dashed blue line in panel a. The black line shows a step function at the radius of the inner contact. The red line shows this profile smeared with the PSF $g(r) = 1/\cosh^2\left(1.76\frac{r}{\sigma}\right)$, where $\sigma = 0.85\ \mu m$. The smeared step function agrees well with the measured data.

## S5. Determining the contact transparency from the measured resistance profile

As discussed in the main text, the two-probe resistance of a Corbino disk in the ballistic regime and with perfectly transmitting contacts, (transmission coefficient $T = 1$), is equal to the Sharvin resistance that corresponds to the radius of its inner contact, $R_{sh}^{in} = \frac{\pi h}{4e^2 K_F(2\pi r_{in})}$. Fig. 2b of the main text presents the radial dependence of resistance, $R(r)$, in the ballistic regime ($T = 6\ K$). Our measurement showed that the overall resistance (equivalent to two probe transport) is $19.5\ \Omega$, somewhat larger than $R_{sh}^{in} \sim 13.67\ \Omega$ at this density. We showed that a small part of this difference appears in the bulk and is due to a finite mean free path ($l_{MR} = 40\ \mu m$) that leads to a small ohmic bulk resistance. However, most of this difference happens at the contacts. For example, from the fitting in Fig. 2c we found that inner contact resistance step height is $0.82 R_{sh}^{in}$, larger than the $0.5 R_{sh}^{in}$ expected for an ideal contact. We will focus here only on the inner contact, because at low temperatures the physics there is simpler than at the outer contact (see section SI11 below). This increased contact resistance reflects a contact transmission that is smaller than one. Following Landauer, we know that a finite transmission leads to a resistance of $\frac{1-T}{T}R_{sh}^{in}$. Adding this to the half Sharvin resistance expected for an ideal contact, we get:

$$R_{contact} = \frac{R_{sh}^{in}}{2} + \frac{1-T}{T}R_{sh}^{in} = R_{sh}^{in}\frac{(2-T)}{2T} \qquad (S5.1)$$

Rearranging the above expression, we get:

$$T = \frac{2}{2R_{contact}/R_{sh}^{in} + 1} \qquad (S5.2)$$

Fig. S5 plots the contact transparency of the inner contact at $T = 6\,K$ as a function of carrier density, obtained by using equation S5.2 and $R_{contact}$ deduced from similar fits as in Fig. 2c, but for the resistance profiles measured at the different densities. We see that over the entire carrier density range the contact transparency is high, on par with the best transparencies achieved with graphene contacts[1,4].

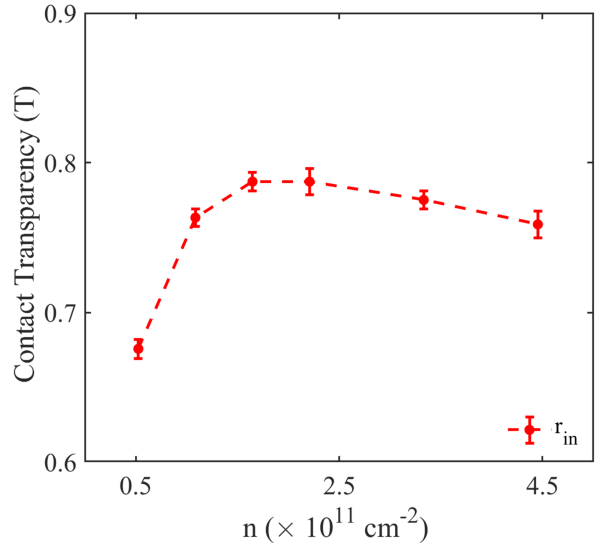

**Fig. S5**: **Density dependence of the inner contact transparency**. We fit $R(r)$ measured at $T = 6\,K$ and different carrier densities and obtain the height of the resistive step at the inner contact (see main text). Using equation S5.2 we obtain the corresponding contact transparency, presented in the figure as a function of the carrier density.

# S6. Determining the momentum relaxing mean free path over the full temperature range.

As explained in the main text, when the temperature is high enough such that the electron flow is hydrodynamic, we can determine the momentum relaxing mean free path, $l_{MR}$, directly from measurements at finite magnetic fields (Fig. 4 in the main text). At low temperature, we can also determine $l_{MR}$ from fitting the imaged resistance profile to Landauer-Sharvin + ohmic dependence + contact resistance, $R_{LS}(r) + R_{ohm}(r) + R_c^{in}(r) + R_c^{out}(r)$ (see Fig. 2c and corresponding text). In this section we use the theoretical expression for the temperature dependence of electron-phonon scattering to interpolate between these measurements and obtain the full temperature dependence of $l_{MR}$. This $l_{MR}$ vs. $T$ curve is then used together with Eq. (1) in the main text to obtain the traces in Fig. 3b.

In general, the momentum relaxing mean free path in graphene is determined by the scattering by disorder and phonons. We will term the former 'impurity scattering', although one has to keep in mind that disorder is not dominated few isolated impurities but is rather by a smooth potential modulation with long spatial scale caused by a distribution of many impurities spaced from the graphene by an hBN spacer. The corresponding disorder and electron-phono mean free paths are $l_{imp}$ and $l_{e-ph}$. The total momentum relaxing mean free path is then given by the Matthiessen sum rule of these two processes:

$$l_{mr} = (l_{imp}^{-1} + l_{e-ph}^{-1})^{-1} \tag{6.1}$$

The density and temperature dependence of the resistance due to phonon scattering is given by[1,5]:

$$\rho_{e-ph}(n,T) = \frac{8D_A^2 k_F}{e^2 \rho_m v_s v_F^2} f_s\left(\frac{\theta_{BG}}{T}\right) \tag{S6.2}$$

where $k_F = \sqrt{\pi n}$ is the Fermi momentum, $D_A$ is the acoustic deformation potential, $\rho_m$ is the mass density of graphene, $v_F$ is the Fermi velocity, $v_s$ is the longitudinal acoustic phonon velocity and $f_s$ is the Bloch Gruneisen function, given by $f_s(z) = \int_0^1 \frac{zx^4\sqrt{1-x^2}\, e^{zx}}{(e^{zx}-1)^2} dx$. The resistivity is translated to the e-ph mean free path using the standard relation valid for graphene:

$$l_{e-ph} = \frac{1}{\rho_{e-ph}(n,T)} \left(\frac{h}{2e^2 k_F}\right) \tag{S6.3}$$

We obtain the density dependence of the impurity mean free path at $T = 6\,K$ directly from our measurements (inset to Fig. 2d in the main text). These measurements are reproduced in blue dots in Fig. S6a together with a polynomial interpolation (blue line). At this temperature, the contribution of electron-phonon scattering is negligible, and thus this curve represents $l_{imp}(n)$. We further assume that the temperature dependence comes entirely from the temperature dependence of the e-ph scattering. If we add to $l_{imp}(n)$ the $l_{e-ph}(n,T)$ from equations (S6.1) - (S6.3) and use the parameters given in Ref[5] ($D_A = 21\ eV, \rho_m = 7.6e-7\ Kgm^{-2}, v_F = 1 \times 10^6\ ms^{-1}$ and $v_s = 2 \times 10^4\ ms^{-1}$ ) we obtain the total $l_{MR}$ at elevated temperatures, shown by the different colored traces in Fig. S6a.

The red and orange dots in Fig. S6a correspond to the $l_{mr}$ obtained from magneto-resistance measurements (as described in the main text) at a density of $n = 4.5 \times 10^{11}\ cm^{-2}$ and $T = 100$ K (orange dot) and at the same density and $T = 140$ K (red dot). We see that these measurements nicely fit the above expression. This can be also seen when we plot $l_{MR}$ vs. temperature (Fig. S6b). The above expression (black line) fits well the three experimentally measured points. This suggest that we can use this expression for obtaining the $l_{MR}$ at temperatures between our low and high temperatures data points.

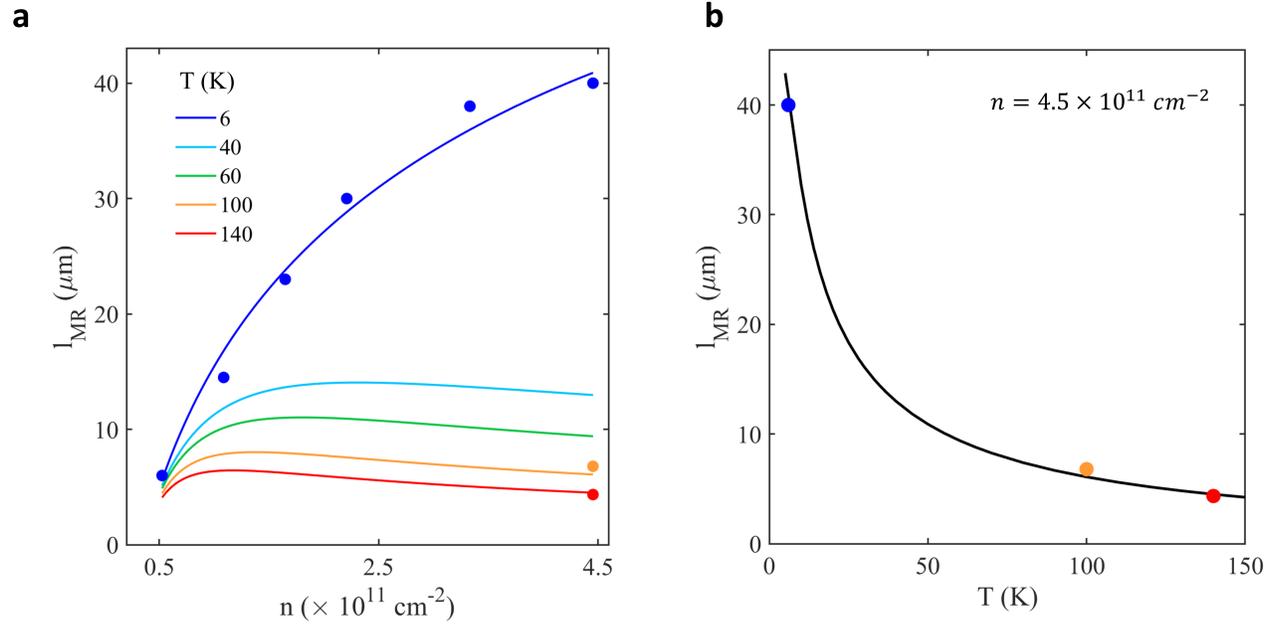

**Fig. S6: $l_{MR}$ vs. temperature and density**. **a.** Measured mean free path as a function of carrier density at $T = 6$ K (blue dots), 100 K (orange dot) and 140 K (red dot) (see text). The blue line is a polynomial fit through the data at 6 K. The lines corresponding to higher temperatures (see legend) include in addition the electron phonon scattering term (see text in this section). **b.** $l_{MR}$ as a function of temperature at $n = 4.5 \times 10^{11} cm^{-2}$. The measured $l_{MR}$ at $T = 6$ K (blue dot) 100 K (orange dot) and 140 K (red dot) are in good agreement with the theoretical expression.

## S7. Additional data at a different carrier density

In the main text, we present data at $V_{BG} = 2\,V$, corresponding to $n = 4.5 \times 10^{11}\,cm^{-2}$. Here, we present additional data from another carrier density ($V_{BG} = 1.5\,V$, $n = 3.3 \times 10^{11}\,cm^{-2}$), demonstrating similar behavior to that presented in the main text.

Fig. S7a shows the imaged $R(r)$ at $T = 6$ K and $n = 3.3 \times 10^{11}\,cm^{-2}$. This curve has similar characteristics as that in Fig. 2b. In Fig. S7b we use a similar fit as in the main text. The dashed line shows a fit to $R_{LS}(r) + R_c^{in}(r) + R_c^{out}(r)$, demonstrating that the measured bulk resistance is predominantly given by the Landauer-Sharvin resistance. Adding the ohmic term $R_{ohm}(r) = R_{sh}^{in} \frac{2r_{in}}{\pi l_{MR}} \log\left(\frac{r}{r_{in}}\right)$ with $l_{mr} = 38\,\mu m$ we obtain the excellent fit shown the dashed red curve.

Fig. S7c shows the bulk resistance $R_{bulk}(r) = R(r) - \left(R_c^{in}(r) + R_c^{out}(r)\right)$, as a function of temperature. We see similar trend to the measurements in the main text: the total bulk resistance first decreased with increasing temperature up to $\sim T = 60$ K and then goes slightly up. Moreover, the spatial dependence within the bulk shows a similar evolution to the one shown in the main text: from a curved resistance profile at low temperatures (when plotted on a logarithmic $r$ axis), $\sim \text{asin}(r_{in}/r)$, to a linear profile, namely $\sim \log(r/r_{in})$, at high temperatures.

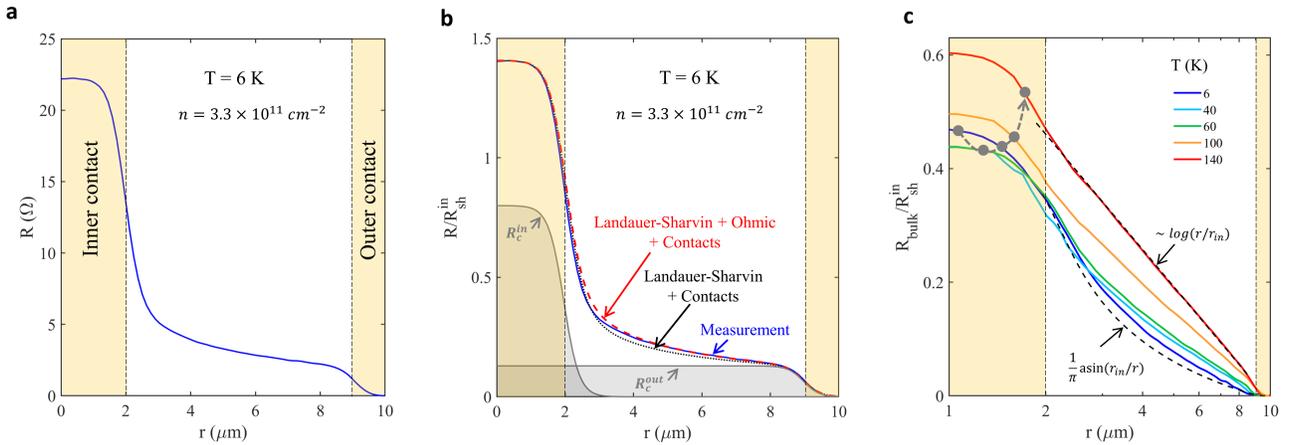

**Fig. S7 | Measured resistance profiles at $n = 3.3 \times 10^{11}\,cm^{-2}$. a.** Measured $R(r)$, at $T = 6\,K$. **b.** Breaking up the resistance to bulk and contact components. The PSF smeared contact resistances, $R_c^{in}(r)$ and $R_c^{out}(r)$ (grey), are as described in the main text. Dotted black line: Fit to $R_{LS}(r) + R_c^{in}(r) + R_c^{out}(r)$, where, $R_{LS}(r) = R_{sh}^{in} \frac{1}{\pi} \text{asin}\left(\frac{r_{in}}{r}\right)$ is the theoretically predicted Landauer-Sharvin bulk geometrical resistance. Red dashed line shows a fit to the same expression with the addition of $R_{ohm}(r) = R_{sh}^{in} \frac{2r_{in}}{\pi l_{MR}} \log\left(\frac{r}{r_{in}}\right)$, with $l_{MR} = 38\,\mu m$. **c.** Measured $R_{bulk}(r) = R(r) - \left(R_c^{in}(r) + R_c^{out}(r)\right)$ at various temperatures, $T$, normalized by $R_{sh}^{in}$ and plotted with a logarithmic $r$ axis.

## S8. Imaging measurements on a second Corbino device

We performed measurements on a second Corbino disk with different dimensions, $r_{in} = 1\ \mu m$ and $r_{out} = 6\ \mu m$. The inset in Fig S8a shows the optical image of this device, fabricated using similar procedure as discussed in SI section 1.

Fig. S8a presents the measured $R(r)$ at $T = 6$ K and $n = 4.5 \times 10^{11}\ cm^{-2}$, where the transport is ballistic. This curve has the same characteristics to that in Fig. 2b. In Fig. S8b we use a similar fit as in the main text. The dashed line shows a fit to $R_{LS}(r) + R_c^{in}(r) + R_c^{out}(r)$, demonstrating that the measured bulk resistance is predominantly given by the Landauer-Sharvin resistance. Adding the ohmic term $R_{ohm}(r) = R_{sh}^{in}\frac{2r_{in}}{\pi l_{MR}}\log\left(\frac{r}{r_{in}}\right)$ with $l_{mr} = 27\ \mu m$ we obtain the excellent fit shown the dashed red curve.

Fig. S7c shows the bulk resistance $R_{bulk}(r) = R(r) - \left(R_c^{in}(r) + R_c^{out}(r)\right)$, as a function of temperature. We see similar trend to the measurements in the main text that: the resistance evolves from a curved resistance profile at low temperatures (when plotted on a logarithmic $r$ axis), $\sim\mathrm{asin}(r_{in}/r)$, to a linear profile, namely $\sim\log(r/r_{in})$, at high temperatures.

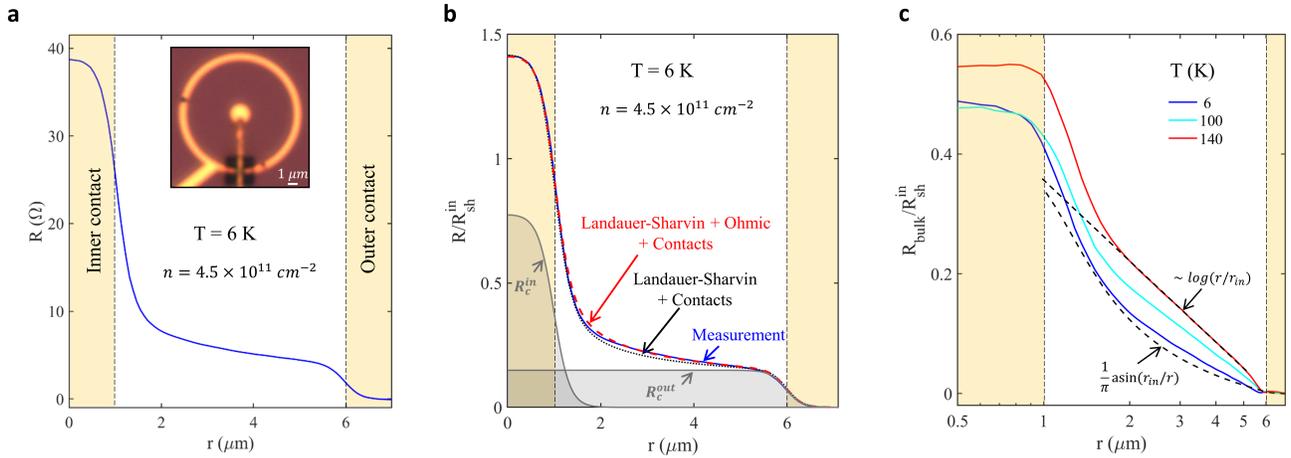

**Fig. S8 | Imaging of a second Corbino device**. **a**. Inset: optical image of the device. Main panel: measured $R(r)$, at $T = 6\ K$. **b.** Breaking up the resistance to bulk and contact components. The PSF smeared contact resistances, $R_c^{in}(r)$ and $R_c^{out}(r)$ (grey), are as described in the main text. Dotted black line: Fit to $R_{LS}(r) + R_c^{in}(r) + R_c^{out}(r)$, where , $R_{LS}(r) = R_{sh}^{in}\frac{1}{\pi}\mathrm{asin}\left(\frac{r_{in}}{r}\right)$ is the theoretically predicted Landauer-Sharvin bulk geometrical resistance. Red dashed line shows a fit to the same expression with the addition of $R_{ohm}(r) = R_{sh}^{in}\frac{2r_{in}}{\pi l_{MR}}\log\left(\frac{r}{r_{in}}\right)$, with $l_{MR} = 27\ \mu m$. **c.** Measured $R_{bulk}(r) = R(r) - \left(R_c^{in}(r) + R_c^{out}(r)\right)$ at various temperatures, $T$, normalized by $R_{sh}^{in}$ and plotted with a logarithmic $r$ axis.

# S9. Derivation of Equation 1 in the main text

In this section we prove the identity:

$$R(r, L_{ee} = (l_{ee}^{-1} + l_{MR}^{-1})^{-1}, L_{MR} = \infty) = R(r, L_{ee} = l_{ee}, L_{MR} = l_{MR}) - \frac{\hbar}{2e^2 k_F l_{MR}} \log(r/r_{in})$$

(s9.1)

which is equation (1) in the main text, used to obtain the clean-limit hydrodynamic flow profile.

In the equation above $R(r, L_{ee}, L_{MR})$ is the bulk resistance profile with momentum-conserving and momentum-relaxing mean free paths of $L_{ee}$ and $L_{MR}$ correspondingly.

The full Boltzmann equation reads as:

$$\vec{v}.\nabla\chi + (l_{ee}^{-1} + l_{MR}^{-1})(\chi - \bar{\chi}) - l_{ee}^{-1} 2\cos(\theta)\overline{\chi \cos(\theta)} = 0 \qquad (S9.2)$$

where $\chi(r,\theta)$ is the Boltzmann function describing the distribution of the carrier's momenta direction, given by $\theta$ at location $r$, and $\bar{\chi}$ is its averaging over $\theta$.

We rewrite equation (S9.2) as,

$$\vec{v}.\nabla\chi + (l_{ee}^{-1} + l_{MR}^{-1})(\chi - \bar{\chi}) - (l_{ee}^{-1} + l_{MR}^{-1}) 2\cos(\theta)\overline{\chi \cos(\theta)} + l_{MR}^{-1} 2\cos(\theta)\overline{\chi \cos(\theta)} = 0 \qquad (S9.3)$$

or,

$$L_0 \chi + l_{MR}^{-1} 2\cos(\theta)\overline{\chi \cos(\theta)} = 0 \qquad (S9.4)$$

where $L_0$ is the Boltzmann equation for a modified problem with $L_{MR} = \infty$ and $L_{ee} = (l_{ee}^{-1} + l_{MR}^{-1})^{-1}$.

Denoting the solution of $L_0$ as $\chi_0$, we look for a solution of the form $\chi = \chi_0 + \delta\chi(r)$.

We find

$$L_0 \delta\chi + l_{MR}^{-1} 2\cos(\phi\theta)\overline{\chi \cos(\theta)} = 0$$

$$\cos(\theta)\partial_r \delta\chi + l_{MR}^{-1} 2\cos(\theta)\overline{\chi \cos(\theta)} = 0$$

$$\partial_r \delta\chi + l_{MR}^{-1} j = 0 \qquad (S9.5)$$

and thus

$$\delta\chi = -l_{MR}^{-1} \frac{I}{2\pi}(\log(r) - \log(r_{in})) \qquad (S9.6)$$

## S10. Boltzmann simulations of interacting flow in a Corbino geometry

To model electron flow through the graphene channels, we employ an approach based on the Boltzmann equation[6–9] that incorporates the effects of both electron-impurity and electron-phonon scattering as well as electron-electron interactions:

$$\partial_t f + \vec{v} \cdot \nabla_{\vec{r}} f = \frac{\partial f}{\partial t}|_{scatt}, \qquad (S10.1)$$

where the scattering integral,

$$\frac{\partial f(\vec{r},\vec{v})}{\partial t}|_{scatt} = -\frac{f(\vec{r},\vec{v}) - n(\vec{r})}{\tau} + \frac{2}{\tau_{ee}} \vec{v} \cdot \vec{j}(\vec{r}), \qquad (S10.2)$$

has two contributions: one from momentum-relaxing scattering, with a rate $\frac{1}{\tau_{MR}}$, and one from momentum-conserving, electron-electron scattering, with a rate $\frac{1}{\tau_{ee}}$. This equation describes the evolution of the semiclassical occupation number $f(\vec{r},\vec{v})$ for a wave packet at position $\vec{r}$ and velocity $\vec{v}$, where $n(\vec{r}) = \langle f \rangle_{\vec{v}}$ is the local charge density, $\vec{j}(\vec{r}) = \langle f\vec{v} \rangle_{\vec{v}}$ the local current density, $\langle ... \rangle_{\vec{v}}$ is the momentum average, and $\frac{1}{\tau} = \frac{1}{\tau_{MR}} + \frac{1}{\tau_{ee}}$. For the sake of simplicity, we consider the case of a circular Fermi surface with $\vec{v} = v_F \hat{\rho}(\theta)$, where $\hat{\rho}$ is the radial unit vector at angle $\theta$. Mean free paths are then simply defined as $l_{MR(ee)} = v_F \cdot \tau_{MR(ee)}$. The term proportional to $\tau_{ee}^{-1}$ is the simplest momentum-conserving scattering term that can be written, assuming that the electrons relax to a Fermi-Dirac distribution shifted by the drift velocity[6,10–12].

The sample is a Corbino disk with inner radius $r_{in}$ and outer radius $r_{out}$. We use polar coordinates in real space as well, with radius $r$ and angle $\phi$. Thanks to a rotational symmetry, the $\phi$ variable drops out of the calculation. Combining everything, the Boltzmann equation takes the form:

$$cos(\theta)\partial_r f - sin(\theta)\frac{1}{r}\partial_\theta f = -\frac{f(r,\theta) - n(r)}{l} + \frac{2}{l_{ee}} \vec{v} \cdot \vec{j}(r)$$

with $= v_F \tau$, $n(r) = \frac{1}{2\pi} \int d\theta\, f(r,\theta)$ and where $j_r(r) = \frac{1}{2\pi} \int d\theta\, f(r,\theta) \cos(\theta)$ is the radial current (the azimuthal current is zero by symmetry).

The rapid transition from a practically-infinite density of states in the metal contact, to the finite density of states at the graphene channel next to it, imposes the following boundary condition:

$$f(r = r_{min}, -\pi/2 \leq \theta \leq \pi/2) = f_{in}$$
$$f(r = r_{max}, \pi/2 \leq \theta \leq 3\pi/2) = 0, \quad \quad \text{(S10.3)}$$

where $f_{in}$ is a constant whose value is set to fix the total current.

The resulting integrodifferential equation is solved numerically using the method of characteristics[13] to invert the differential part of the equation, and an iterative method to solve the integral part.

Based on the solution for $f$, one finds the total current as $I = 2\pi j_r$ and the electrochemical potential as the electron density $n(r)$ divided by the density of states at the Fermi level.

Further, the contact resistance can be deduced by assuming the following form for $f$ at the two contacts: $f(r = r_{min} - \varepsilon, \theta) = f_{in}$ at the inner contact, and $f(r = r_{out} + \varepsilon, \theta) = 0$ at the outer contact, where $\varepsilon$ is infinitesimal.

# S11. Temperature dependence of the outer contact resistance

The experiments described in Fig. 3b of the main text shows that as the temperature increases, a small step gradually builds up near the outer contact. We recall that in all the curves in that figure (corresponding to measurement temperatures from $T = 6\,K$ to $140\,K$) we subtracted the **same** contact step functions $R_c^{in}(r)$ and $R_c^{out}(r)$, those obtained from fitting the resistance profile at $T = 6\,K$ (Fig. 2c). This means that the contact steps that we see in Fig. 3b reflect the **difference** between the contact resistance at finite temperatures and that at $T = 6\,K$. The fact that we observe a finite outer contact step at high $T$ therefore suggests that the outer contact resistance is larger at higher $T$.

In our Boltzmann numerical calculations (Fig. 3c) we observed a similar effect, but even stronger. There upon decreasing of $l_{ee}$ (which corresponds to increasing $T$ in the experiment) we see the buildup of a resistance step at the outer contact. We note that we use a similar procedure to the one used for the experimental curves - subtracting from all the Boltzmann curves the same contact resistance steps, the one obtained in the calculation with $l_{ee} = \infty$. The buildup of contact resistance step with decreasing $l_{ee}$ in Fig. 3c therefore implies that the outer contact resistance increases with decreasing $l_{ee}$.

The Boltzmann calculation allows us to identify that this phenomenon originates from the transition between highly non-local ballistic flow at $l_{ee} = \infty$ and a locally equilibrated hydrodynamic flow at small $l_{ee}$. In a ballistic flow, the angular distribution of carriers in graphene, just outside the inner contact, has half of the angles populated by hot carriers emitted from the inner contact. Going toward the outer contact, the hot electrons get collimated to a smaller angular spread (the simple analogy would be the angle distribution of light rays reaching from the sun to an observer. As the observer gets further away from the sun, the distribution of light rays (/hot electrons) becomes more collimated). At a radius $r$ the hot electrons are collimated to an angular spread of $\Delta\theta = 2\,\text{asin}\left(\frac{r_{in}}{r}\right)$. As we have shown in the paper, the corresponding $r$ dependence of the resistance is $R(r) = R_{sh}^{in}\frac{1}{\pi}\text{asin}\left(\frac{r_{in}}{r}\right)$. For $r = r_{out} \gg r_{in}$ the collimated beam is very narrow, $\Delta\theta \approx 2\frac{r_{in}}{r_{out}}$. The outer contact resistance can be readily obtained from the value of the resistance function at $r_{out}$, since at the outer contact itself it equals zero. Namely, the theoretically predicted outer contact resistance in the ballistic regime is:

(ballistic) $$R_{contact}^{out} = R_{sh}^{in}\frac{1}{\pi}\text{asin}\left(\frac{r_{in}}{r_{out}}\right) \approx R_{sh}^{in}\frac{1}{\pi}\frac{r_{in}}{r_{out}} \tag{S11.1}$$

Indeed, this is the value that we obtain in the Boltzmann calculations in the ballistic regime ($l_{ee} = \infty$).

When we examine the Boltzmann calculations deep in the hydrodynamic case ($l_{ee} \ll r_{in}, r_{out}, r_{out} - r_{in}$) we see that we obtain a different outer contact resistance. In fact, in this case we observe that this resistance roughly equals the resistance one would obtain in the diffusive regime, namely, half the Sharvin resistance corresponding to $r_{out}$, $R_{sh}^{out} = \frac{\pi h}{4e^2} \frac{1}{k_F(2\pi r_{out})}$. Consequently,

(hydrodynamic) $$R_{contact}^{out} = \frac{1}{2} R_{sh}^{out} = R_{sh}^{in} \frac{1}{2} \frac{r_{in}}{r_{out}}$$ (S11.2)

At the same time, in the hydrodynamic regime, the angular distribution of the electrons near the outer contact is $\sim \cos(\theta)$. Comparing equations (11.1) and (11.2) we see that $R_{contact}^{out}$ in the hydrodynamic regime is larger by a factor $\pi/2$ than that in the ballistic regime (we ignore here higher orders in $l_{ee}$). This corresponds to the increase of outer contact resistance with decreasing $l_{ee}$ seen in Fig. 3c. This factor reflects the fact that the current density carried by a highly collimated angular distribution is larger by a factor of $\pi/2$ than the current density carried by a $\cos(\theta)$ distribution.

In the experiment we observe similar effect to that in the Boltzmann numerics, although the increase of the outer contact resistance with temperature is somewhat smaller. We note also that this feature is approaching the noise level of our experiment. It is likely that the reduced amplitude of the effect in the experiment, compared to the ideal theory, reflects the fact that the collimation effect described above is much more sensitive to experimental imperfections, such as roughness of the outer contact, more than the other effects reported in this paper that display very good agreement with the theory.